\documentstyle[12pt,epsf]{article}
\font\sf=cmss10                    

\makeatletter
\@addtoreset{equation}{section}

\makeatother

\def\del{\partial}
\def\Dslash{\not{\hbox{\kern-4pt $D$}}}
\def\dslash{\not{\hbox{\kern-2pt $\del$}}}
\def\gslash{\not{\hbox{\kern-2pt $\gamma$}}}
\def\g5{\gamma_5}

\def\ra{\rightarrow}

\newcommand{\mathbold}[1]{\mbox{\boldmath $\bf#1$}}
\def\mJ{\mathbold{J}}

\def\ma{\mathbold{a}}
\def\mb{\mathbold{b}}
\def\mc{\mathbold{c}}

\def\bbbz {{\sf Z\!\!\!Z}}
\def\sl2z{SL(2,\bbbz)}
\def\fracs#1#2{\textstyle\frac #1#2}
\def\gcd{{\cal GCD}}

\begin{document}
\newcommand{\onefigure}[2]{\begin{figure}[htbp]
\begin{center}\leavevmode\epsfbox{#1.eps}\end{center}\caption{#2\label{#1}}
         \end{figure}}
\newcommand{\figref}[1]{Fig.~\protect\ref{#1}}
\setlength{\parskip}{1.35ex}
\setlength{\parindent}{0em}

\thispagestyle{empty}
{\flushright{\small MIT-CTP-2747\\hep-th/9805220\\}}

\vspace{.3in}
\begin{center}\Large {\bf Constraints On The BPS Spectrum Of 
${\cal N}$= 2, D=4 Theories With A-D-E Flavor Symmetry}
\end{center} 

\vspace{.3in}
\begin{center}
{\large Oliver DeWolfe, Tam\'as Hauer, Amer Iqbal
and Barton Zwiebach}

\vspace{.4in}
{ {\it Center for Theoretical Physics,\\
Laboratory for Nuclear Science,\\
Department of Physics\\
Massachusetts Institute of Technology\\
Cambridge, Massachusetts 02139, U.S.A.}}
\vspace{.2in}

E-mail: {\tt odewolfe,hauer,iqbal@ctp.mit.edu,
zwiebach@irene.mit.edu}
\end{center}
\begin{center}May 1998\end{center}

\vspace{0.1in}
\begin{abstract}
BPS states of ${\cal N}=2, D=4$ Super Yang-Mills theories with $ADE$
flavor symmetry arise as junctions joining a D3-brane to a set of
7-branes defining the enhanced flavor algebra. We show that the
familiar BPS spectrum of SU(2) theories with $N_f \leq 4$ is simply
given by the set of junctions whose self-intersection is bounded below
as required by supersymmetry.  This constraint, together with the
relations between junction and weight lattices, is used to establish the
appearance of arbitrarily large flavor representations for the case of
$D_{n\geq 5}$ and $E_n$ symmetries.  Such representations are required
by consistency with decoupling down to smaller flavor symmetries.
\end{abstract}

\newpage

\section{Introduction}

It has become quite clear that in IIB superstring theory $(p,q)$
strings \cite{witten} and string junctions \cite{aharony,
schwarzreview} are actually on the same footing.  More precisely,
states defined by a fixed set of charges on a moduli space of
backgrounds are sometimes realized as strings, but more generally, as
junctions. This viewpoint has been obtained by examination of
backgrounds with 7-branes \cite{GZ, mgthbz, hauer}, and with 7-branes
together with 3-branes \cite{bergmanfay,spalinski}.  We focus here on
the case of a single 3-brane on the background of some 7-branes
defining an $ADE$ type configuration.  The theory on the 3-brane is a
four dimensional ${\cal N}=2$ Yang-Mills theory and when the 3-brane
sits away from the 7-branes, the gauge symmetry of the 7-branes
appears as a flavor symmetry for the 3-brane. BPS states of the
3-brane field theory are now recognized as junctions joining the
3-brane and (some of) the 7-branes.

In a recent interesting work Mikhailov, Nekrasov and Sethi
\cite{nekrasov} discussed a constraint that selected the junctions
giving the BPS spectrum for the SU(2) theory with no matter.  This
constraint is based on a subtle argument comparing pairs of
junctions. Motivated by this work, we show here that a very simple
constraint also gives the well-known BPS spectrum, not only for the
pure SU(2) theory discussed in \cite{nekrasov} but also for the cases
when we have $N_f \leq 4$ flavors.  This constraint requires any BPS
junction $\mJ$ with asymptotic charges $(p,q)$ to have
self-intersection
\begin{equation}
(\mJ, \mJ) - \gcd(p,q) \geq  -2,
\label{intro_eq}
\end{equation}
and it arises as follows.  By construction
\cite{nekrasov,dewolfezwiebach}, $(\mJ , \mJ) = \# (J\cdot J)$ where
$J$ is the two-dimensional cycle associated to $\mJ$ in the F/M theory
picture and $\# $ denotes intersection number.  In order for $\mJ$ to
be BPS, $J$ must be holomorphic, and then $ \# (J\cdot J) = 2g-2 +b$,
where $g$ is the genus of the curve and $b$ is the number of
boundaries.  We show that $b$ can be identified with $\gcd(p,q)$, the
greatest common divisor of the charges, and since $g\geq 0$,
(\ref{intro_eq}) 
follows. We then explore this
constraint on the BPS spectrum for the general ADE case.

For the cases of $D_{n\geq 5}$ and $E_6, E_7 $ and $E_8$ flavor groups
our discussion is based on the recently-found relation between the
junction lattice and the corresponding Lie-algebra weight vector
lattice \cite{dewolfezwiebach}.  The self-intersection of a junction
has two contributions, a negative-definite one given by minus the
length squared of the corresponding weight vector, and a
positive-definite one from a quadratic form on the asymptotic $(p,q)$
charges of the junction.  These charges are seen on the D3-brane as
the electric and magnetic charges of the BPS state.  In contrast to
the case of $D_{n \leq 4}$ where only vectors, spinors and singlets
appear, here the constraint permits arbitrarily large representations
for sufficiently large $(p,q)$ charges.  Encouraged by the precise
agreement between our predictions and the known spectra, one can
speculate that all states allowed by (\ref{intro_eq}) are present in
the spectra of the $D_{n\geq 5}$, $E_6, E_7 $ and $E_8$ theories as
well.  In support of this, we show that consistency with smaller
algebras after a brane decoupling requires some of the large
representations to belong to the spectrum. However other
representations decouple completely and so their presence in the BPS
spectrum cannot be confirmed by these consistency arguments.  Finally
we examine the remnant of the $\sl2z$ duality group that acts on the
BPS spectra of the various theories.

\section{Selection rule based on Self-Intersection}
\label{s:selectionrule}

$(p,q)$ strings in Type IIB string theory compactified on a manifold
$B$ are holomorphic curves of F/M theory compactified on a four real
dimensional elliptically fibered manifold $X$ with base $B$. These
curves are formed by taking a geodesic on $B$ and a cycle of the
elliptic fiber above the geodesic. $[p,q]$ 7-branes  on the base
$B$ correspond to singular fibers of $X$. A D3-brane lifts to a regular
elliptic fiber $F_{0}$ above  its position  on $B$.

In ref.\cite{bds}, $\cal N$=2 $SU(2)$ Seiberg-Witten theory was
interpreted as the worldvolume theory of a D3-brane in the presence of
mutually non-local 7-branes. Charged states in the D3-brane theory are
junctions with legs on 7-branes and ending on the D3-brane with
non-vanishing asymptotic charge. In the F/M theory picture BPS states
of charge $(p,q)$ correspond to holomorphic curves with a $(p,q)$
cycle of $F_{0}$ as the boundary.

The manifold $X$ is hyper-K\"ahler of vanishing first Chern class and
therefore has a two-sphere worth of complex structures \cite{vgs}.
The elliptic fiber $F_{0}$ corresponding to the D3-brane is
holomorphic in one of the complex structures.  The BPS states are
curves holomorphic in a complex structure orthogonal to that of the
elliptic fiber. Thus the space of allowed complex structures for the
curves corresponding to BPS states is a circle.

The self-intersection number of a smooth holomorphic curve $J$ of
genus $g$ with $b$ boundary components in a complex surface $X$ is
equal to the degree of the normal bundle of $J$ in $X$
\cite{Hartshorne,vgs}, therefore
\begin{eqnarray}
\label{mainprev}
\#(J\cdot J) = degN_{J/X}=
\int_{J}c_{1}(N_{J/X})
=-\chi(J)=2g-2+b.
\end{eqnarray}
since the first Chern class of $X$ is zero. In the case of a single
D3-brane the boundary of $J$ is a $(p,q)$ cycle of the elliptic fiber
$F_{0}$.  If $p$ and $q$ are not relatively prime, the greatest common
divisor $\gcd(p,q) \geq 1$ gives the number $b$ of boundary components
(at the D3-brane), this is necessary for $J$ to be smooth. Consider
now the junction $\mJ$ associated to $J$, which by construction
\cite{nekrasov, dewolfezwiebach} satisfies $(\mJ, \mJ) = \# ( J \cdot
J )$.  Since the genus $g$ is nonnegative, it follows from
(\ref{mainprev}) that
\begin{eqnarray}
(\mJ,\mJ)-\gcd(p,q) \geq -2\, .
\label{theformula}
\end{eqnarray}
This constraint will be the primary tool in our analysis.

The selection rule of \cite{nekrasov} was based on the fact that
submanifolds holomorphic in the same complex structure have positive
intersection number. By intersecting the junction shown in
\figref{amer}(a) with an $(r,0)$ string starting at the 7-brane, it
was shown that the junction may only be BPS if $-r^{2}+mr \geq 0$. We
recover this as follows.  Imagining that the strings end on
3-branes (to make a well-defined junction) we have at least two
boundary components, and therefore $(\mJ,\mJ)\geq 0$. On the other
hand, from the rules of \cite{dewolfezwiebach} we find $(\mJ,\mJ) =
-r^2 + rm$, thus reproducing the claimed result.

\onefigure{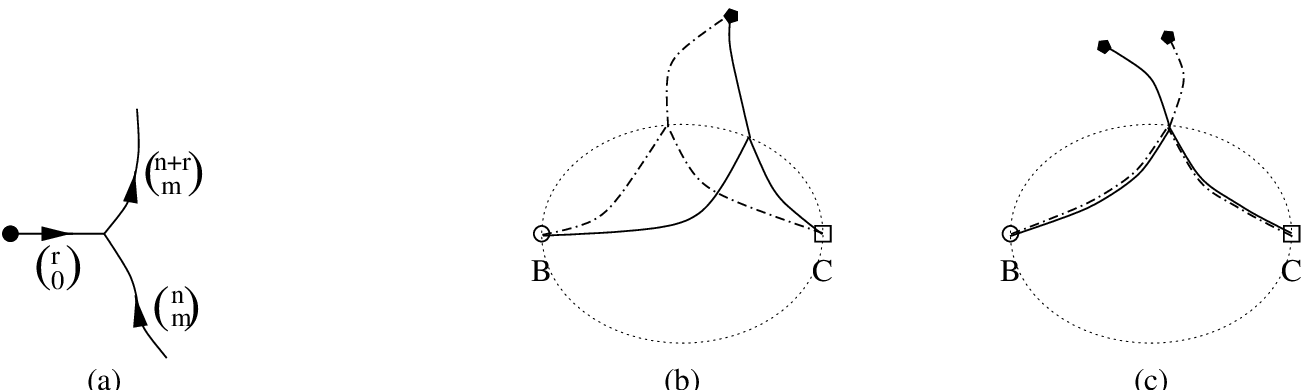}{(a) A junction with one prong on a 7-brane which may
not be supersymmetric for certain values of the charges. (b) Two BPS
junctions of the SW theory are not holomorphic in the same complex
structure. (c) Two junctions made holomorphic in the same complex
structure represent BPS states of two different D3-brane theories.}

The fact that holomorphic curves intersect positively was also used to
select the BPS spectrum of ${\cal N}=2$ $ SU(2)$ pure SW
\cite{nekrasov}.  Cycles corresponding to BPS states of different
charges ending on the same D3-brane are holomorphic in different
complex structures (\figref{amer}(b)).  This is the case because,
having different charges, the associated junctions must depart the
3-brane in different directions, and thus hit the curve of
marginal stability at different points. This requires identical prongs
departing the 7-branes to do so at different angles, a signal of
different complex structures.  The cycles can be made holomorphic in
the same complex structure but in this case they have boundaries on
different elliptic fibers of $X$ (\figref{amer}(c)). These two curves
can be considered BPS states of two different D3-brane worldvolume
theories. The spectrum does not change as long as a D3-brane is
outside the curve of marginal stability, therefore if there is a BPS
state of charge $(p,q)$ in one theory it must also exist in the other.
It was shown that under the assumption that a state with magnetic
charge one exits, the only states which have positive intersection
number with this state are the ones with magnetic charge $\pm$1 or
$0$. It is known from semi-classical analysis that in the weak
coupling regime the only BPS states are the dyons with magnetic charge
$\pm 1$ and the W-boson. Thus this argument gives exactly the known
spectrum if the initial assumption about the existence of a BPS state
of magnetic charge one is correct.

A direct argument based on self-intersection number does not require
comparing states of different theories. Consider the state of charge
$(Q_B+ Q_C,Q_C - Q_B)$ in the ${\cal N} =2$ $SU(2)$ SW theory and
represented by a junction $\mJ,$ with $Q_B$ legs on the B brane and
$Q_C$ legs on the C brane. In the notation of \cite{dewolfezwiebach}
\begin{eqnarray}
\mJ= Q_B\, \mb+ Q_C \,\mc\,.
\end{eqnarray}
Using the rules given in \cite{dewolfezwiebach} we can calculate the
self-intersection number.
\begin{eqnarray}
\#(J\cdot J)=(\mJ,\mJ) =-(Q_B- Q_C)^{2} \geq -1\,,
\end{eqnarray}
implying that the magnetic charge $(Q_C - Q_B)$ of a BPS state is either
$0$ or $\pm 1$, thus recovering the familiar result.

\section{Recovering the familiar BPS spectra}
\label{s:recoveringfamiliar}

\newcommand{\be}{\begin{equation}}
\newcommand{\ee}{\end{equation}}
\newcommand{\bea}{\begin{eqnarray}}
\newcommand{\eea}{\end{eqnarray}}
\newcommand{\nn}{\nonumber}
\newcommand{\bean}{\begin{eqnarray*}}
\newcommand{\eean}{\end{eqnarray*}}
\newcommand{\integ}[2]{\int\limits_{#1}^{#2}\!\!}
\newcommand{\ordo}[1]{{\cal O}(#1)}
\newcommand{\myref}[1]{(\ref{#1})}
\newcommand{\secref}[1]{sec.~\protect\ref{#1}}
\newcommand{\unit}{1}
\newcommand{\mat}[4]{\left(\begin{array}{cc} #1 & #2 \\ #3 & #4
\end{array}\right)}
\newcommand{\vect}[2]{({#1\atop
#2})}
\newcommand{\mpl}{\!+\!}
\newcommand{\mmi}{\!-\!}
\newcommand{\spl}{\!\!+\!\!}
\newcommand{\smi}{\!\!-\!\!}

This section is devoted to testing the selection rule proposed above
by applying it to brane configurations of familiar field theories with
known spectrum. The examples will be the well known Seiberg-Witten
theories \cite{seibergwitten1,seibergwitten2} with $N_f=0,1,2,3,4$
flavors. The pure ${\cal N}=2$ SYM theory is realized on a D3-brane in
the vicinity of two mutually nonlocal 7-branes with charges $[1,1]$
and $[-1,1]$ ({\bf B}- and {\bf C}-brane) which stand for the two
strong-coupling singularities on the ``u-plane''. The theories with
quarks are obtained by adding up to four $[1,0]$ 7-branes ({\bf
A}-branes). These conventions agree with \cite{mgthbz} and differ by
an overall SL(2,Z) transformation by $({1 1\atop 0 1})$ from the one
used in \cite{seibergwitten2}.


A state in the D3-brane field
theory is represented by a junction which ends on the 3-brane
and on some or all of the 7-branes and is characterized by the
invariant charges
($Q_C,Q_B,Q_A^1\ldots Q_A^{N_f}$) of each brane.
The self-intersection number and the central charges of such a state
are determined as \cite{dewolfezwiebach}:
\bea
({\bf J},{\bf J}) &=& -(Q_C-Q_B)^2+
(Q_C\mmi Q_B)(Q_A^1\mpl\ldots\mpl Q_A^{N_f})\mmi
((Q_A^1)^2\mpl\ldots+(Q_A^{N_f})^2)
\label{SWint} \\
(p,q)&=&(Q_C+Q_B+Q_A^1\mpl\ldots\mpl Q_A^{N_f},Q_C\mmi Q_B).  \eea It
is straightforward to find all the solutions to \myref{theformula} in
terms of the invariant charges. The complete list of states for the
theories with $N_f=1 \ldots 4$ is provided in the following tables. In
the last columns we list the representations of the flavor symmetry
algebra, $so(2N_f)$.  
\bea
\begin{array}{ll}
\begin{array}[t]{||l|r||}
\hline
(Q_C,Q_B) \;\;\; N_f=0& (p,q) \\ \hline\hline
(1,1) & (2,0) \\ \hline
(n\spl 1,n) & (2n\spl 1,1) \\ \hline
\end{array}
&
\begin{array}[t]{||l|r|c||}
\hline
(Q_C,Q_B,Q_A) \;\;\; N_f=1& (p,q) & so(2) \\ \hline\hline
(1,1,0) & (2,0) & 0 \\ \hline
(0,0,1) & (1,0) & 1 \\
(1,1,-1) & (1,0) & -1 \\ \hline
(n\spl 1,n,0) & (2n\spl 1,1) & 1/2 \\
(n,n\smi 1,1) & (2n,1) & -1/2 \\ \hline
\end{array}
\end{array} 
\nn\eea
\bea
\begin{array}[t]{||l|r|c||}
\hline
\begin{array}{l}(Q_C,Q_B,Q_A^1,Q_A^2)\hspace{.5in} N_f=2\end{array}&
  \begin{array}{r}(p,q)\end{array}&so(4) \\
  \hline\hline
  \begin{array}{l}(1,1,0,0)\end{array} &
  \begin{array}{r}(2,0)\end{array} & ({\bf1},{\bf1}) \\ \hline
  \begin{array}{ll}(0,0,1,0),&(0,0,0,1) \\ (1,1,-1,0),&(1,1,0,-1)
    \end{array} & \begin{array}{r}(1,0)\\(1,0)
                  \end{array}    & ({\bf2},{\bf2}) \\ \hline
  \begin{array}{ll}(n\spl 1,n,0,0),& (n,n\smi 1,1,1) \end{array} &
  \begin{array}{r}(2n\spl 1,1)\end{array} &
                                   ({\bf2},{\bf1}) \\ \hline
  \begin{array}{l}(n,n\smi 1,1,0),(n,n\smi 1,0,1)\end{array} &
  \begin{array}{r}(2n,1)\end{array} & ({\bf1},{\bf2}) \\ \hline
\end{array}
\nn
\eea

\bea
\begin{array}{c}
\begin{array}[t]{||l|r|c||}
\hline
  \begin{array}{l}(Q_C,Q_B,Q_A^1,Q_A^2,Q_A^3)\hspace{1.5in} N_f=3
     \end{array}&
  \begin{array}{r}(p,q)\end{array}&so(6) \\
  \hline\hline
  \begin{array}{l}(1,1,0,0,0)\end{array} &
  \begin{array}{r}(2,0)\end{array} & {\bf1} \\ \hline
  \begin{array}{lll}(0,0,1,0,0),&(0,0,0,1,0),&(0,0,0,0,1) \\
                    (1,1,-1,0,0),&(1,1,0,-1,0),&(1,1,0,0,-1)
    \end{array} & \begin{array}{r}(1,0)\\(1,0)
                    \end{array}    & {\bf6} \\ \hline
  \begin{array}{l}(n\spl 1,n,0,0,0),
                     (n,n\smi 1,1,1,0),(n,n\smi 1,1,0,1),
                     (n,n\smi 1,0,1,1)
    \end{array} & \begin{array}{r}(2n\spl 1,1)
                  \end{array}    & {\bf4} \\ \hline
  \begin{array}{l}(n\smi 1,n\smi 2,1,1,1),
                     (n,n\smi 1,1,0,0),(n,n\smi 1,0,1,0),
                     (n,n\smi 1,0,0,1)
                     \end{array} &
  \begin{array}{r}(2n,1)\end{array} & {\bf\overline{4}} \\ \hline
  \begin{array}{l}(n,n\smi 2,1,1,1)\end{array} &
  \begin{array}{r}(2n\spl 1,2)\end{array} & {\bf1} \\ \hline
\end{array}
\end{array} \nn 
\eea 
\bea
\begin{array}{c}
\begin{array}[t]{||l|r|c||}
\hline
\begin{array}{l}(Q_C,Q_B,Q_A^1,Q_A^2,Q_A^3,Q_A^4)\hspace{1.3in}
   N_f=4\end{array}&
\begin{array}{r}(p,q)\end{array}&so(8) \\
\hline\hline
\begin{array}{l}
(n\smi m,n\smi 3m,m,m,m,m)
\end{array} &
\begin{array}{r}(2n,2m)\end{array} & {\bf1} \\ \hline
\begin{array}{l}
  (n\smi m,n\smi 3m,m,m,m,m\spl 1),\ldots \\
  (n\smi m\spl 1,n\smi 3m\spl 1,m,m,m,m\smi 1),\ldots \end{array} &
\begin{array}{r}(2n\spl 1,2m)\end{array}    & {\bf8_v} \\ \hline
\begin{array}{l}
  (n\smi m,n\smi 3m\smi 1,m,m,m,m\spl 1),\ldots \\
  (n\smi m\spl 1,n\smi 3m,m,m,m,m\smi 1),\ldots \end{array} &
\begin{array}{r}(2n,2m\spl 1)\end{array}    & {\bf8_s} \\ \hline
\begin{array}{l}
  (n\smi m\spl 1,n\smi 3m,m,m,m,m),
  (n\smi m,n\smi 3m\spl 1,m,m,m,m)  \\
  (n\smi m,n\smi 3m\smi 1,m,m,m\spl 1,m\spl 1),\ldots \end{array} &
\begin{array}{r}(2n\spl 1,2m\spl 1)\end{array}    & {\bf8_c} \\ \hline
\end{array}
\end{array} \nn
\eea 
For the $N_f < 4$ tables $n$ is an arbitrary integer, and there is an
identical set of states (not listed) with all invariant charges
reversed and all representations conjugated.  For $N_f=4$ in the first
row, $n$ and $m$ are coprime while in the rest of the table $p$ and
$q$ must be coprime; ``$\ldots$'' stands for obvious permutations of
the {\bf A}-brane invariant charges.  The $N_f=4$ results
for $n,m=0$ were anticipated by \cite{imamura}.

Instead of using the invariant charges, we may also solve directly for
the allowed representations using the results of
\cite{dewolfezwiebach}. Consider the case of $so(8)$ and note
that the four conjugacy classes are identified by $(p,q)$ mod 2.
After translating the invariant charges to weight vectors we obtain
$(\mJ, \mJ) = -\vec\lambda \cdot \vec\lambda$. Then \myref{SWint}
becomes: 
\bea
\vec\lambda \cdot \vec\lambda \leq 2 - \gcd(p, q).  
\eea 
This implies that the singlet representation survives with
$\gcd(p,q)=2$ and the {\bf 8}'s with $\gcd(p,q)=1$. This is in
accord with the fact that there are two null junctions (having
zero intersection with any other junction) of charges $(2,0)$ and
$(0,2)$ in the $so(8)$ lattice which, when added to another junction,
do not change its weight vector. Therefore if a representation is
allowed, it is present with every $(p,q)$ compatible with its
conjugacy class.

Notice that the listings give exactly the weak-coupling spectrum of
the Seiberg-Witten theories with $N_f\le4$. We emphasize that the
above table is simply the full set of solutions to \myref{theformula},
no extra assumption about the field theories was made. Thus we find
that the supersymmetric string states which are allowed by our simple
selection rule are {\em in one-to-one correspondence} with the states
of the spectrum of both the $N_f=0\ldots 3$
\cite{seibergwitten2,ferraribilal} and the $N_f=4$
\cite{seibergwitten2,ferrari} models. This result tempts us to leave
the area of models with known spectrum and propose \myref{theformula}
as a tool to investigate those field theories which we have little
information about. This is what we shall do in the next section.

\section{Larger flavor symmetries and decoupling}
\label{s:largersymmetries}

Configurations of parallel seven-branes can produce any Lie algebra of
$ADE$ type.  We consider here the worldvolume theory on a three-brane
in the presence of $D_{n\geq 5}$ and $E_6$, $E_7$, $E_8$ backgrounds,
and constrain the BPS spectra of these theories using the
self-intersection criterion.  In terms of its associated Lie algebra
weight vector $\vec\lambda$ and charges $(p,q)$, the self-intersection
of a junction $\mJ$ can be written \cite{dewolfezwiebach}
\begin{eqnarray}
\label{ncons}
 (\mJ, \mJ) = -\vec\lambda \cdot \vec\lambda + f(p,q)
\geq -2 + \gcd(p, q) \label{Jsquared}\,.
\end{eqnarray}
where $f(p,q)$ is a quadratic form of definite sign given for the
various algebras in Table \ref{forms}.

\begin{table}[ht]
\begin{center}
\def\st{\vrule height3ex depth 2ex width0pt}
\begin{tabular}{|c|c|c|} \hline \hline
Algebra & $f(p,q)$ & $\mu^2(p,q)$\\ \hline
$A_n$ & $-\fracs1n \, p^2$ & --- \\
$D_n$ & $\fracs14 q^2 (n-4)$ & --- \\
$E_6$ & $\fracs13 p^2 - pq + q^2$ & $p^2 - pq + q^2$\\
$E_7$ & $\fracs12 p^2 - 2pq + \fracs52 q^2$ & $p^2 + q^2$\\
$E_8$ & $p^2 - 5 pq + 7 q^2$ & $p^2 - pq + q^2$\\ \hline
\end{tabular}
\end{center}
\caption{The quadratic forms $f(p,q)$ and rescaled mass per unit length
squared $\mu^2(p,q)$ for the $D_n, n \geq 5$ and $E_6$, $E_7$, $E_8$
algebras.\label{forms}}
\end{table}

For the $A_n$ singularities obtained by collapsing $n+1$ {\bf
A}-branes the only consistent junctions fall into fundamentals,
antifundamentals and singlets of $A_n$. We will not consider this case
any further.  For $D_{n \geq 5}$ and $E_6$, $E_7$, and $E_8$, $f(p,q)$
is positive-definite instead of negative-definite or vanishing.
Arbitrarily large representations, having arbitrarily large
weight vectors, can then be associated to junctions satisfying (\ref{ncons})
if sufficiently large $(p,q)$ values are chosen.  As we will show,
consistency with brane decoupling transitions actually requires
arbitrarily large representations.  Brane decoupling has been considered
in other contexts in \cite{imamura}. Also indicated in the table is
$\mu^2 \equiv |p-q\, \tau|^2$, the rescaled mass per unit length
squared of a junction \cite{schwarz}, evaluated for constant
$\tau$ values \cite{dasguptamukhi}. Even though arbitrarily large
representations appear, only a finite number of representations yield
states with mass-squared less than or equal to any fixed value.

As a 7-brane is moved to infinity, the only junctions that survive are
those having zero invariant charge associated to the brane; all others
become infinitely massive and decouple.  Thus, as the brane is moved
away, the invariant charges, the asymptotic charges $(p,q)$, and the
self-intersection number of the surviving junctions do not change.
Junctions satisfying the self-intersection constraint before the
removal of the brane continue to do so afterwards.

The decoupling of a single brane induces the removal of a simple root
$\vec\alpha_i$ with dual weight $\vec\omega^i$.  The rank decreases by
one, and one can define the $u(1)$ generator $H^* \propto \vec\omega^i
\cdot \vec{H}$, where $\vec{H}$ are the Cartan generators of the
parent algebra. $H^*$ commutes with the full subalgebra, and its
eigenvalue on weight vectors is denoted by $Q^*$. For example, for
$so(10) \ra so(8)$, the brane $\ma_1$ is decoupled, removing the
simple root $\vec\alpha_1$.  Fixing the normalization, we act on a
weight vector $\vec\lambda$ to find
\begin{eqnarray}
\label{qstar}
Q^* &=& H^* (\vec \lambda) =
2 \vec\omega^1 \cdot \vec  \lambda = 2 (A^{1i} \vec\alpha_i) \cdot (a_j
\,\vec\omega^j) = 2\,A^{1i} a_i
\nonumber \\
 &=& 2a_1 + 2a_2 + 2a_3 + a_4 + a_5 = 2\,Q_A^1 + Q_B - Q_C \\
&=& 2\, Q_A^1 - q\,, \nonumber
\end{eqnarray}
where $A^{ij}$ is the inverse Cartan matrix of $so(10)$ and use was
made of the relation between Dynkin labels and invariant charges
(\cite{dewolfezwiebach}, eqn.~(6.27)).  Thus an $so(10)$ junction
survives to $so(8)$ if
\begin{eqnarray}
Q_A^1 = \fracs12 (Q^* + q) = 0 \quad \ra \quad Q^* = -q \,.  \label{decouple}
\end{eqnarray}
Since $H^*$ commutes with $so(8)$, all states in a given $so(8)$
representation have the same value of $Q^*$. Depending on the $q$
value of the original $so(10)$ representation, the $so(8)$
representation will either decouple or survive as a whole. The
analysis of brane removal for other symmetries follows along the above
lines. The results are summarized in Table \ref{transitions}, where we
show the brane that decouples, the simple root that is removed, the
coefficients of the Dynkin labels in the expression for $Q^*$, and
the invariant charge on the brane to be removed. Note that this charge
is written entirely in terms of the charges $(p,q)$ and $Q^*$.

\begin{table}
\begin{center}
\begin{tabular}{|c|c|c|c|c|} \hline
Transition & Brane & Root & $u(1)$ charge $Q^*$ & Invariant charge on
removed brane \\
\hline
$so(10) \ra so(8)$ & $\ma_1$ & $\vec\alpha_1$ & $(2,2,2,1,1)$  & $Q^1_A =
\fracs12 (Q^* + q)$ \\
$E_6 \ra so(10)$ & $\mc_2$ & $\vec\alpha_5$ & $(2,4,6,5,4,3)$  & $Q^2_C =
\fracs13 (2p - 3q - Q^*)$  \\
$E_7 \ra E_6$ & $\ma_6$ & $\vec\alpha_6$ & $(2,4,6,5,4,3,3)$ & $Q_A^6  =
\fracs12 (3q - p - Q^*)$ \\
$E_7 \ra so(12)$ & $\mc_1$ & $\vec\alpha_1$ & $(2,3,4,3,2,1,2)$ & $Q_C^1 =
Q^* + p - 2q$ \\
$E_8 \ra E_7$ & $\ma_7$ & $\vec\alpha_7$ & $(2,4,6,5,4,3,2,3)$& $Q_A^7  =
3q - p - Q^*$ \\
$E_8 \ra so(14)$ & $\mc_1$ & $\vec\alpha_1$ & $(4,7,10,8,6,4,2,5)$  &
$Q_C^1 = Q^* + 2p - 5q$ \\ \hline
\end{tabular}
\end{center}
\caption{Various brane decoupling transitions, the brane and simple root
removed, the $u(1)$ charge $Q^*$ and the invariant charge on the removed
brane in terms of $Q^*$, $p$ and $q$.\label{transitions}}
\end{table}

We now consider in further detail the case of $so(10) \ra so(8)$.  We
will show that to get the complete spectrum for $so(8)$ with all
possible $(p,q)$ charges, we need arbitrarily large representations in
$so(10)$. Indeed, consider the decompositions
\begin{eqnarray}
{\bf 10} &\ra& ({\bf 8_v})_0 + {\bf 1}_2 + {\bf 1}_{-2} \,, \nonumber \\
{\bf 16} &\ra& ({\bf 8_c})_1 + ({\bf 8_s})_{-1} \,, \\
{\bf \overline{16}} &\ra& ({\bf 8_s})_1  + ({\bf 8_c})_{-1} \,, \nonumber
\end{eqnarray}
where the subscript is $Q^*$.  Since by (\ref{decouple}) only
junctions with charge $q = -Q^*$ survive in $so(8)$, these $so(10)$
representations only produce an ${\bf 8_v}$ with $q=0$ and ${\bf
8_s},{\bf 8_c}$ with $q =\pm 1$.  ${\bf 8}$'s with larger $q^2$ are
embedded in other $so(10)$ representations.

We now show that an ${\bf 8}$ or ${\bf 1}$ of $so(8)$ with fixed
$(p,q)$ charges arises from a unique representation of $so(10)$. Let
$R$ denote an $so(10)$ representation that contains an ${\bf
8_v}$. This will be the case if $R$ contains the weight $\vec\lambda_k
= (k,1,0,0,0)$ for some integer $k$. It follows from (\ref{qstar})
that $\vec\lambda_k$ has $Q^* = 2k+2$.  Moreover, $\vec\lambda_k \cdot
\vec\lambda_k = k^2 + 2k + 2$, and since the $(p,q)$ charges are
coprime (\ref{Jsquared}) becomes
\begin{eqnarray}
\label{Jso10}
-\vec\lambda \cdot \vec\lambda + \fracs14 q^2 = (\mJ, \mJ) \geq -1 \,
\quad 
\to \quad
q^2 \geq (2k+2)^2\,.
\end{eqnarray}
For $\vec\lambda_k$, as well as the rest of the states giving the
${\bf 8_v}$, to survive decoupling we must have $q = - Q^* = -
(2k+2)$. This fixes $k$ in terms of $q$, and as a consequence
$\vec\lambda_k$ is also fixed. Note that the ${\bf 8_v}$ junctions
saturate the self-intersection bound.  Therefore, $\vec\lambda_k$ must
be one of the longest weights in $R$; any longer weight would violate
(\ref{Jso10}). There is a unique representation which contains a given
weight and none longer, and hence $R$ is unique. Since the longest
weights in a representation occur with multiplicity one, the ${\bf
8_v}$ occurs in $R$ only once.  Analogous arguments apply for ${\bf
8_s}$, ${\bf 8_c}$ and ${\bf 1}$; they are also embedded uniquely in
$so(10)$ representations.  We indicate the $so(10)$
representation that contains each $so(8)$ representation for given
values of $(p,q)$ in Table \ref{so10spec}.

\begin{table}
\begin{center}
\begin{tabular}{|c|c|c|c|} \hline
$so(8)$ & $(p,q)$ & Range of $m$ & $so(10)$ highest weight \\ 
\hline \hline
${\bf 1}$ & $(2n,2m)$ & All $m$ &$(|m|,0,0,0,0)$ \\ \hline
${\bf 8_v}$& $(2n+1, 2m)$  & $|m| \geq 1$ & $(|m|-1,1,0,0,0)$ \\
& &  $m = 0$ & $(1,0,0,0,0)$ \\            \hline
${\bf 8_s}$ & $(2n, 2m+1)$ & $m \geq 0$ & $(m,0,0,0,1)$ \\
            &           & $m \leq -1$ & $(|m|-1,0,0,1,0)$ \\ \hline
${\bf 8_c}$ & $(2n+1,2m+1)$ & $m \geq 0$ & $(m,0,0,1,0)$ \\
            &           & $m \leq -1$ & $(|m|-1,0,0,0,1)$ \\
\hline
\end{tabular}
\end{center}
\caption{$so(8)$ representations of various $(p,q)$ charges
embedded in $so(10)$.\label{so10spec}}
\end{table}

Thus we have shown that when an additional brane is brought in from
infinity to an $so(8)$ configuration, the ${\bf 8}$ and ${\bf 1}$
representations of various $q$ charges transform in arbitrarily large
representations of $so(10)$; thus consistency requires all these
states to be in the $so(10)$ spectrum.  Note however that there are
junctions in $so(10)$ representations which decouple completely for
any values of $q$; these are the ones which do not include ${\bf 8}$
or ${\bf 1}$ representations in their decomposition.  For example, the
${\bf 126}$ of $so(10)$ has highest weight $\vec\lambda_0 =
(0,0,0,0,2)$ and decomposes as
\begin{eqnarray}
{\bf 126} \ra ({\bf 56_v})_0 + ({\bf 35_c})_2 + ({\bf 35_s})_{-2} \,.
\end{eqnarray}
However, $\vec\lambda_0 \cdot \vec\lambda_0 = 5$, and so by
(\ref{Jso10}) $q^2 \geq 16$; there is no acceptable value of $q$ for
which $q = - Q^*$.  Such representations are permitted by
self-intersection for appropriate values of $q$, but consistency with
$so(8)$ makes no statement about their presence in the $so(10)$
spectrum.

In Table \ref{embed} we show how $so(8)$ representations with specific
$(p,q)$ charges are embedded in successively larger groups.  Table
\ref{ellipse} presents junctions with various $(p,q)$ charges and
representations for the smallest possible values of the quadratic form
$f(p,q)$ for $E_6$.  Conjugacy requires that the ${\bf 27}$ and ${\bf
351}$ representations occur only for $p=1$ (mod 3), the ${\bf
\overline{27}}$ and ${\bf \overline{351}}$ occur only for $p=2$ (mod 3),
and the {\bf 1}, {\bf 78} and {\bf 650} occur for $p=0$ (mod 3). The
presence of the representations marked with a dagger is not required
by consistency with decoupling.

\begin{table}
\begin{center}
\begin{tabular}{|c|c|c|c|c|c|} \hline
$(p,q)$ & $so(8)$ & $so(10)$ & $E_6$ & $E_7$ & $E_8$ \\ \hline
$(1,0)$ & ${\bf 8_v}$ & ${\bf 10}$ & ${\bf 27}$ & ${\bf 56}$ & ${\bf
248}$ \\ 
$(0,1)$ & ${\bf 8_s}$ & ${\bf 16}$ & ${\bf 78}$ & ${\bf 912}$ & ${\bf
147\,250}$ \\
$(1,1)$ & ${\bf 8_c}$ & ${\bf \overline{16}}$ & ${\bf 27}$ & ${\bf
133}$ & 
${\bf 3875}$ \\
$(2,0)$ & ${\bf 1}$ & ${\bf 1}$ & ${\bf \overline{27}}$ & ${\bf 133}$ &
${\bf 3875}$ \\
$(1,2)$ & ${\bf 8_v}$ & ${\bf 45}$ & ${\bf 351}$ & ${\bf 27664}$ & ${\bf
6\,899\,079\,264}$ \\ \hline
\end{tabular}
\end{center}
\caption{$so(8)$ representations of given $(p,q)$ charges
embedded in successively larger groups.\label{embed}}
\end{table}

\begin{table}
\begin{center}
\begin{tabular}{|c|c|c|} \hline
$f(p,q)$ &  Possible $\pm (p,q)$ &  Representations \\ \hline
$1/3$ & $  (1,0)$, $  (1,1)$, $  (2,1)$  & ${\bf 27}$, ${\bf
\overline{27}}$  \\ 
$1$ & $  (0,1)$, $  (3,2)$, $  (3,1)$ & ${\bf 1}^{\dagger}$, ${\bf78}$
\\ 
$4/3$& $  (2,0)$, $  (2,2)$, $  (4,2)$& ${\bf 27}$, ${\bf
\overline{27}}$ \\ 
$7/3$ & $  (1,2)$, $(1,-1)$,  $(4,3)$, $  (4,1)$, $  (5,3)$, $  (5,2)$
& ${\bf 27}^{\dagger}$, ${\bf \overline{27}}^{\dagger}$, ${\bf 351}$,
${\bf \overline{351}}$ \\  
$3$ & $(3,0)$, $(3,3)$, $(6,3)$ &  ${\bf 1}^{\dagger}$,
${\bf78}^{\dagger}$, ${\bf 650}$  \\ \hline 
\end{tabular}
\end{center}
\caption{Junctions with various $(p,q)$ charges and representations
realizing 
the smallest values of the quadratic form $f(p,q)$ for
$E_6$. \label{ellipse}} 
\end{table}

\section{Duality constraints}

The theory with $so(8)$ flavor symmetry has an $\sl2z$ duality acting
on the BPS charges, which induces an action on the representations via
the $S_3$ permutation group implementing $so(8)$ triality
\cite{seibergwitten2}.  Given a representation $R$ with highest weight
vector $(a_1, a_2 , a_3 ,a_4)$ an element of $g\in \sl2z$ will map it
to a representation where the labels $a_1, a_3 $ and $a_4$ have been
permuted according to the element of $S_3$ associated to $g$ by the
homomorphism $h: \sl2z \to S_3$. More concretely, the spectrum of the
theory, defined by the BPS charges and representations $\sum_i\{ (p_i,
q_i) ; R_i \}$, is invariant under the action of a group with elements
of the form $(g, h(g))$, where $g\in \sl2z$ acts on the $(p,q)$
charges, and $h(g)\in S_3$ acts on the representation $R_i$. This
claim is readily verified by examination of the table corresponding to
$N_f=4$ in section 3.

We will now develop corresponding results for $so(10), E_6, E_7$ and
$E_8$. While in the $so(8)$ case the duality conjecture is well
supported by additional evidence, in the other cases such evidence is
not available. Our analysis is therefore in essence a proposal for the
duality symmetry of the spectrum of these unfamiliar theories.  This
proposed symmetry of the spectrum implies nontrivial constraints on
representations and multiplicities, especially for the representations
that are not required by the decoupling argument discussed in the
previous section.  The analysis has one new element: we claim that in
each case the relevant $\sl2z$ transformations must preserve the
quadratic form $f(p,q)$.  (For $so(8)$ the quadratic form vanishes
\cite{dewolfezwiebach}, and therefore the full $\sl2z$ is relevant.)
The symmetry transformation must relate representations of the same
size with weight vectors of equal lengths.  Thus if the quadratic form
$f(p,q)$ is not left invariant, one could find that $\sl2z$ action
maps junctions allowed by self-intersection to forbidden
junctions.\footnote{One could, in principle, allow a larger subgroup
of $\sl2z$ and demand that representations that can become forbidden
junctions be eliminated. It seems, however, that such a constraint
would actually eliminate representations that we know must be
present.}  We will find that in general the group $\sl2z$ is broken
down to a subgroup $M$. We will then find a homomorphism $h: M\to
{\cal A}$ to the outer automorphisms of the particular algebra.  The
symmetry group of the spectrum will be generated by elements of the
form $(m, h(m))$ with $m\in M$ and $h(m) \in {\cal A}$.

For the case of $so(10)$ the group $\bbbz_2$ of automorphisms of this
algebra is generated by $\sigma: a_4 \leftrightarrow a_5$. 
Conjugacy classes of representations are given by $C= 2a_1 +
2a_3 +a_4 - a_5$ (mod 4).  One can readily see that $\sigma$ exchanges
representations with $C=1$ and $C=3$, while leaving invariant
those with $C=0,2$. The conjugacy class is correlated with
asymptotic charges as $C= 2p-q$ (mod 4) \cite{dewolfezwiebach}. For
$so(10)$ the group $\sl2z$ is broken down to the subgroup $M$ of
transformations preserving the quadratic form $q^2$.  The elements of
$M$ are given by
\begin{equation}
M_\pm^n  = \pmatrix {\pm 1 & n\cr 0 & \pm 1} \,, \quad n \in \bbbz\,.
\end{equation}
By checking the action of the matrices on $(p,q)$ and thus on $C$, one
readily verifies that the homomorphism $M \to \bbbz_2$ is defined by
$\{ M_+^{odd}, M_-^{even} \} \to \sigma$, while all others map to the
identity. The group of symmetries is therefore given by $\{
(M_+^{even}, e) , (M_-^{odd},e), (M_+^{odd}, \sigma ) ,( M_-^{even},
\sigma) \}$. This is an infinite abelian group. It is simple to verify
that the necessary spectrum of $so(10)$ listed in Table \ref{so10spec}
is consistent with this group of symmetries.  The same is true for the
$so(6)$ case ($N_f=3$) discussed in section 3.

For the case of $E_6$ the group $\bbbz_2$ of automorphisms is
generated by the element $\sigma: (a_1,a_2) \leftrightarrow
(a_4,a_5)$. Conjugacy classes of representations are given by $C= a_1
- a_2 +a_4 - a_5$ (mod 3), and $\sigma$ exchanges representations with
$C=1$ and $C=2$, while leaving invariant ones with $C=0$. We also have
$C= p$ (mod 3) \cite{dewolfezwiebach}.  In this case the group $\sl2z$
is broken down to the subgroup $M(6)$ of transformations preserving
the quadratic form given in Table \ref{forms}.  The elements of
$M(6)$ are given by
\begin{equation}
M^0_\pm(6)  = \pmatrix {\pm 1 & 0\cr 0 & \pm 1}\,, \quad
M^1_\pm(6)  = \pmatrix {\pm 1 & \mp 3 \cr \pm 1 & \mp 2 }\,, \quad
M^2_\pm(6) = \pmatrix {\mp 2 & \pm 3 \cr \mp 1 & \pm 1}\,.
\end{equation}
By checking the action of the matrices on $(p,q)$ and thus on $C$, one
readily verifies that the homomorphism $M \to \bbbz_2$ is defined by
$M_+^i(6) \to e$, $M_-^i(6) \to \sigma$ for $i=1,2,3$.  The group
of symmetries is therefore given by $\{ (M_+^i, e) ,( M^i_-(6),
\sigma) \}$. This is isomorphic to $\bbbz_2 \times
\bbbz_3$.

For the case of $E_7$ the group of automorphisms is trivial. Conjugacy
classes of representations are given by $C= a_3 + a_4 +a_6 + a_7$ (mod
2), and $C= p+q$ (mod 2) \cite{dewolfezwiebach}.  In this case the
group $\sl2z$ is broken down to the subgroup $M(7)$ of transformations
preserving the quadratic form given in the table.  The elements of
$M(7)$ are given by
\begin{equation}
M^0_\pm (7) = \pmatrix {\pm 1 & 0\cr 0 & \pm 1}\,, \quad
M^1_\pm (7) = \pmatrix {\pm 2 & \mp 5 \cr \pm 1 & \mp 2 }\,.
\end{equation}
Since there is no algebra automorphism, we expect that the above
matrices act on $(p,q)$ leaving $C$ invariant. This is readily
verified to be the case. The group of symmetries is simply $\{
(M^i_\pm(7) , e) \}$. This is isomorphic to $\bbbz_4$.

For the case of $E_8$ the group of automorphisms is trivial and there
only a single conjugacy class.  In this case the group $\sl2z$ is
broken down to the subgroup $M(8)$ of transformations preserving the
relevant quadratic form.  The elements of $M(8)$ are given by
\begin{equation}
M^0_\pm (8) = \pmatrix {\pm 1 & 0\cr 0 & \pm 1}\,, \quad
M^1_\pm (8) = \pmatrix {\pm 2 & \mp 7 \cr \pm 1 & \mp 3 }\,, \quad
M^2_\pm (8)= \pmatrix {\pm 3 & \mp 7 \cr \pm 1 & \mp 2}\,.
\end{equation}
Since there is no algebra automorphism, nor conjugacy classes, the
action of $M(8)$ on the spectrum must leave the $E_8$ representations
invariant. The group of symmetries is simply $\{ (M^i_\pm(8) , e) \}$.
This is isomorphic to $\bbbz_2 \times \bbbz_3$.

\section{Conclusions and open questions}\label{s:Conclusion}

We have considered configurations of 7-branes with $D_n$ and $E_n$
symmetry giving rise, on a 3-brane probe, to four-dimensional
${\cal N} = 2$ theories with global $D_n$ and $E_n$ symmetry
respectively. The theories with global exceptional symmetry, in
particular, are only defined for strong coupling; they are believed to
be non-Lagrangian interacting fixed-point theories \cite{seibergfive,
minahan}, and little is known about them.

We have shown that the self-intersection constraint selects the
junctions giving the well-known spectrum for the familiar ${\cal N}
=2$ SU(2) SYM theories with $N_f= 0, \ldots 4$. This striking result
led us to investigate the $D_{n\geq 5}$ and $E_n$ theories as well.
This constraint, together with the results of \cite{dewolfezwiebach},
allowed us to derive some new facts about their BPS spectra.  In fact,
we suspect that the junctions allowed by the self-intersection
constraint are all BPS and all appear in the spectrum.  More work will
be necessary to be sure about this.

We have exhibited a major change in the nature of the BPS spectrum
when we go from the familiar theories to the case of $D_{n\geq 5}$ and
$E_n$ flavor symmetries.  In the latter cases arbitrarily large
representations are required, while in the former only a few
representations of the flavor group appear.  While for the familiar
theories all BPS states arise from junctions of self-intersection
minus one or zero, in the less familiar theories all self-intersection
numbers are realized, and in general the junctions correspond to
curves of higher genus. We have also carried out a preliminary
investigation of duality constraints on the BPS spectrum. These
constraints relate representations and their multiplicities for
different values of the asymptotic charges.

While we believe to have made some concrete progress in elucidating
the BPS spectrum of the mysterious theories, much remains to be
investigated.  The multiplicities of representations not constrained
by decoupling are not known.  The representations of supersymmetry
associated to general BPS states are also unknown.  These are
questions that are related to the quantization of zero modes of
general junctions, and have been addressed for particular situations
in \cite{callan,OB}. A rich spectrum of states is suggested also by
the authors of \cite{vafawarner} in 6D theories. A complete
description of the BPS spectrum of four-dimensional theories with $ADE$
flavor symmetries appears to be within reach.

\subsection*{Acknowledgments}

We would like to thank O.~Aharony, O.~Bergman and A.~Matusis for
useful discussions, and A.~Hanany and J.~Minahan for helpful
correspondence.  This work was supported by the U.S.\ Department of
Energy under contract \#DE-FC02-94ER40818.

\end{document}